\documentclass[a4paper]{jpconf}
\usepackage{graphicx}
\newcommand{\prd}{Phys. Rev. D}
\newcommand{\prl}{Phys. Rev. Lett.}
\newcommand{\pr}{Phys. Rev.}
\newcommand{\plb}{Phys. Lett. B}

\newcommand{\epja}{Euro. Phys. J. A}  
\newcommand{\npa}{Nucl. Phys. A}
\newcommand{\ap}{Ann. Phys.}
\newcommand{\ie}{{\it i.e., }}

\begin{document}
\title{Dense matter in strong magnetic fields}

\author{Monika Sinha}

\address{Institute for Theoretical
Physics, J.~W.~Goethe-University, D-60438 Frankfurt-Main, Germany\\
Indian Institute of Technology Rajasthan, Jodhpur 342011, Rajasthan, India
}

\begin{abstract}
Compact stars having strong magnetic fields (magnetars) have 
been observationally determined to have surface magnetic fields 
of order of $10^{14}-10^{15}$ G, the implied internal field 
strength being several orders larger. We study the equation 
of state and composition of hypernuclear matter and quark matter 
- two forms of dense matter in strong magnetic fields.
We find that the magnetic field has substantial influence on 
the properties of hypernuclear matter and quark matter for 
magnetic field $B \ge 10^{17}$~G and $B \ge 10^{18}$~G respectively. 
In particular 
the matter properties become anisotropic. Moreover, above a
critical field $B_{cr}$, both hypernuclear and quark matter 
show instability, although the values of $B_{cr}$ are different
for two kinds of matter.
\end{abstract}

\section{Introduction}

Soft $\gamma$-ray repeaters and anomalous X-ray pulsars are commonly
believed to be magnetars - the compact stars with surface magnetic
fields $B_s\sim 10^{14}-10^{15}$ G.
The interpretation of astrophysical observations of magnetars requires
good knowledge of the properties of dense matter in the presence of
large magnetic fields. The properties of dense matter inside the
compact objects is poorly known due to lack of the proper knowledge of
strong interactions that are relevant at densities of interest. As a
consequence, many phenomenological models of dense matter have been
proposed over the years.  The models of dense matter can tentatively
be divided into two broad classes: one class includes matter made of
ordinary hadronic matter; the second class is deconfined quark-gluon
state with about equal number of up, down and strange quarks, known as
the strange quark matter (SQM). In both cases the underlying
constituent particles are fermions which are interacting via exchange
of bosons. Fermions in strong magnetic field experience two well-known
quantum mechanical effects: the Pauli paramagnetism and the Landau
diamagnetism. The first is due to the interaction of the spin of the
fermion with the magnetic field and therefore, is relevant for both
charged and uncharged fermions. The second effect is relevant only for
charged fermions, and is particularly strong for light particles,
which in the case of compact stars are the leptons.

Since one can not exclude the possibility of a density dependent field
profile which is favorable for local magnetostatic equilibrium inside
a star, we consider the hyperonic matter under the influence of
density dependent magnetic field.  Among many effective
phenomenological models to study the quark matter, the MIT bag
model \cite{1974PhRvD...9.3471C} and the Nambu-Jona-Lasinio (NJL)
model \cite{1961PhRv..124..246N,1961PhRv..122..345N} are the most
popular ones.  However, the NJL model does not account for confinement
property of QCD, while the bag model cannot account for the chiral
symmetry breaking, although it is built in a manner as to confine
through an {\it ad hoc} bag pressure. We will base our
discussion of quark matter in strong magnetic field and at nonzero
temperature on a model originally introduced by Dey {\it et. al.}
\cite{1998PhLB..438..123D}. In this model quark masses are density dependent
ensuring chiral symmetry restoration at high density and the quarks
interact among themselves via the Richardson potential (RP)
\cite{1979PhLB...82..272R} in which asymptotic freedom and
confinement are built in. In the next section we discuss the general
effect of magnetic field on fermionic matter. In Sec. \ref{res} we
discuss the results concerning the equation of state (EoS) of hypernuclear
and quark matter in strong magnetic fields. Our conclusions are 
presented in Sec. \ref{conc}.

\section{Magnetized matter}

In the presence of a magnetic field the motion of charged fermions is
Landau quantized in the plane perpendicular to the direction of the
magnetic field. If the field direction is assumed to be the direction
of $z$-axis the single particle energy of a charged particle with mass
$m$ in the $n$-th Landau level is $\epsilon_n = \sqrt{p_z^2 + m^2 + 2
  n e |Q|{B}}$, $p_z$ being the momentum parallel to the field (which
is in the $z$ direction of Cartesian system of coordinates)  and $Q$
being the charge of the particle in units of proton charge. Evidently,
the momentum of a particle in the $x$-$y$ plane is quantized.  
The phase space sampling for such a particle is therefore modified
according to the rule (the spin degeneracy is included)
\begin{equation}
2\int_0^{p_F} d^3p \longrightarrow e|Q|B \sum_{n=0}^{n_{max}} (2-\delta_{n,0})
\int_{-p_{F,n}}^{p_{F,n}} dp_z \int_0^{2\pi} d\phi, 
\end{equation}
where $\phi$ is the azimuthal angle and $p_{F,n} = \sqrt{p_F^2-2ne|Q|B}$
with $p_F$ being the Fermi momentum.

In the presence of electromagnetic field the energy momentum tensor of
a fermionic system in the rest frame of matter is $T^{\mu \nu} =
T^{\mu \nu}_m + T^{\mu \nu}_f$, where the matter part of energy
momentum tensor is
\begin{equation}
\label{eq:Tmatter}
T^{\mu\nu}_m = \left[\begin{array}{cccc} 
                    \varepsilon_m & 0 & 0 & 0\\
                    0 & P_m-MB & 0 & 0\\
                    0 & 0 & P_m-MB & 0\\
                    0 & 0 & 0 & P_m 
               \end{array}\right],
\end{equation}
and in the absence of electric field the field part of the energy
momentum tensor is
\begin{equation}
\label{eq:Tfield}
T^{\mu\nu}_f = \frac{B^2}{8\pi} \left[\begin{array}{cccc} 
                    1 & 0 & 0 & 0\\
                    0 & 1 & 0 & 0\\
                    0 & 0 & 1& 0\\
                    0 & 0 & 0 &-1 
               \end{array}\right].
\end{equation}
The matter energy density $\varepsilon_m$ is obtained by integrating
the single particle energies of each species over appropriate phase
space volume and then adding their contributions. The thermodynamic pressure
at zero temperature is $P_m=\sum_i\,\mu_i n_i - \varepsilon_m$, where
$\mu$ and $n$ are chemical potential and number density of
corresponding species.  The total energy density of the system is
given by the sum of the matter and field contributions $\varepsilon =
\varepsilon_m + {B^2}/{8\pi}$.  It is clear from
Eqs. ~(\ref{eq:Tmatter}) and ~(\ref{eq:Tfield}) that the presence of
magnetic field makes the pressure anisotropic: the pressure in
the perpendicular to the magnetic field direction is $P_\perp = P_m
- MB + {B^2}/{8\pi}$, whereas the pressure in the direction parallel
to the magnetic field is $P_\parallel = P_m - {B^2}/{8\pi}$.

We study the hyperonic matter in the presence of strong magnetic
field within the nonlinear Boguta-Bodmer-Walecka model 
\cite{1974AnPhy..83..491W,1987ZPhyA.327..295G}. The model is
described in detail in Ref. \cite{2013NuPhA.898...43S}. The
field profile has been adopted according to \cite{1997PhRvL..79.2176B}
\begin{equation}
\label{profile}
B\left(\frac{n_b}{n_0}\right)=B_s+B_c\left\{1-\exp\left[{-\beta \left(
\frac {n_b}{n_0} \right)^\gamma}\right]\right\}.
\end{equation}
The parameters $\beta$ and $\gamma$ control the relaxation 
from the central value $B_c$ to the asymptotic value at the 
suface $B_s$. Here $n_b$ is the baryon number density, $n_0$
is the normal nuclear matter density.

In the case of quark matter at finite temperature, we consider matter
 composed of  $u$, $d$ and $s$ quarks which interact via 
the Richardson potential~\cite{1979PhLB...82..272R}
\begin{equation}
\label{richpot}
V(q^2) =- \frac49  ~\frac \pi {{\rm ln}[1+(q^2+m_g^2)/\Lambda^2]} \frac{1}{(q^2+m_g^2)},
\end{equation}
where $m_g$ is gluon mass and $\Lambda$ is a scale parameter.
The finite gluon mass is responsible for screening in medium 
and is related to the screening length $D$ via
\begin{equation}
 m_g^2 = D^{-2} = \frac{2\alpha_0}{\pi}\sum_{i=u,d,s} k_F^i \mu_i^*,
\end{equation}
where $\alpha_0$ is the perturbative quark gluon coupling,
$\mu_i^* = \sqrt{(k^i_F)^2 + m_i^2}$ is the Fermi energy (the 
chemical potential at zero temperature), $k^i_F$ is the Fermi
momentum and $m_i$ the quark mass.
In this model the quark masses vary with density as
\cite{1998PhLB..438..123D}
\begin{equation}
\label{eq:masses}
m_i = M_{i} + M_q\, {\rm sech} \left(\nu \frac{n_b}{n_0}
\right), \qquad i=u,d,s,
\end{equation}
where $\nu$ is a parameter.  At
large $n_b$
the quark mass $m_i$ falls off from its constituent value
$M_q$ to its current value $M_{i}$.

The kinetic part of the energy density for a particular
 quark flavor in the presence of magnetic field and at non-zero
 temperature is given by
\begin{equation}
 \varepsilon_{kin} = \frac 3{(2\pi)^3} e|Q|B
\sum_{n=0}^\infty (2-\delta_{n,0})\int_0^{2\pi}
d\phi \int_{-\infty}^\infty f(\epsilon) \epsilon\, \, \, dk_z,
\end{equation}
where $f(\epsilon)$ is Fermi distribution function.
The potential part of the energy density due to interaction
between the flavors $i$ and $j$ is given by
\begin{eqnarray}
\varepsilon_{pot}^{ij}
&=& \frac{e^2 |Q_i||Q_j|}{(2\pi)^5}
B^2\sum_{n_i} \sum_{n_j} (2-\delta_{n_i,0}) (2-\delta_{n_j,0})
\nonumber \\
 && \hspace{-1cm}\int_0^{2\pi}\! \! \! d\phi_i \int_0^{2\pi}\!\!   \!  d\phi_j
\int_{-\infty}^\infty  \!  \! \! dk_z^i
 \int_{-\infty}^\infty  \!  \! \! dk_z^j f(\epsilon_i) f(\epsilon_j)NV(q^2)S ,
\end{eqnarray}
where
\begin{eqnarray}
 N = \frac{(\epsilon_i+m_i)(\epsilon_j+m_j)}{4\epsilon_i\epsilon_j},\,\,\,\,\,
S = 1+ \frac{k_i^2 k_j^2}{(\epsilon_i+m_i)^2 
(\epsilon_j+m_j)^2} + \frac{2{\mathbf k_i}\
cdot{\mathbf k_j}}{(\epsilon_i+m_i)(\epsilon_j+m_j)}.\nonumber
\end{eqnarray}
The matter pressure is then given by 
\begin{equation}
\label{eq:pressure}
P_m = \sum_i \mu_i n_i + Ts -\varepsilon, \quad
\end{equation}
where $T$ is the temperature, $s$ is the entropy density and $\varepsilon$ 
is the sum of the kinetic and potential energies. 
For further details see Ref. \cite{2013PhRvD..88b5008S}.

\section{Results}\label{res}

\begin{figure}
\begin{center}
\includegraphics[width=7cm,height=6cm]{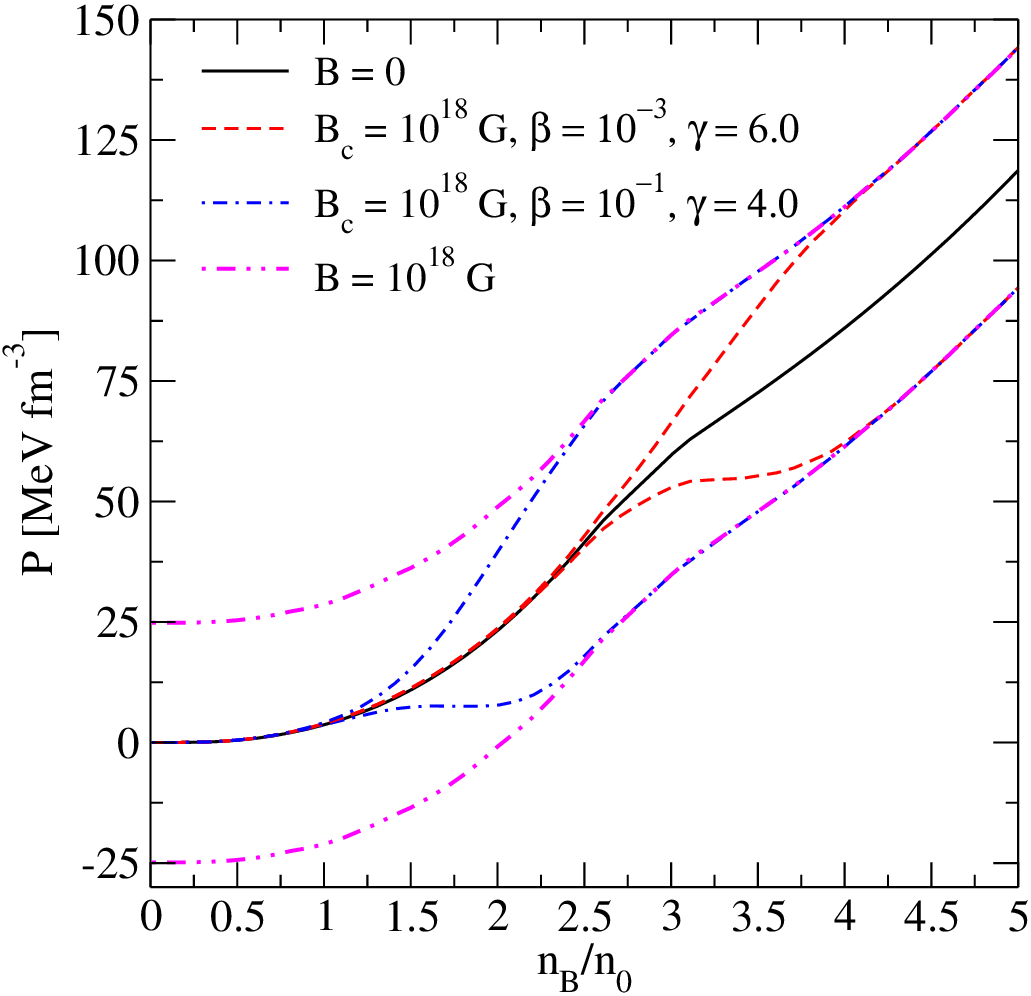}
\hskip 1.0cm
\includegraphics[width=7cm,height=6cm]{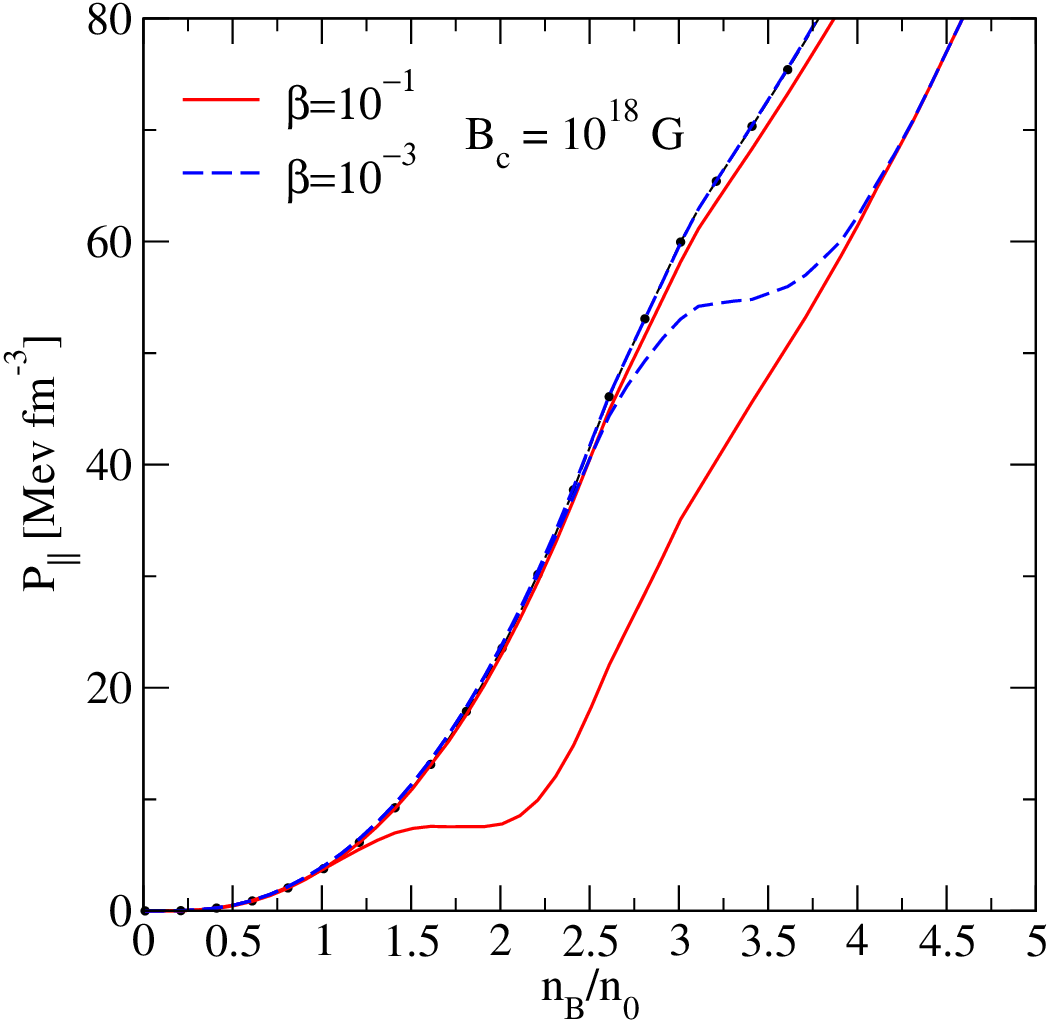}
\caption{ {\it Left panel} (a): Variation of total pressure as a function
  of normalized baryon number density for fixed magnetic fields $B_c =
  0$ (solid line) and $B_c = 10^{18}$~G with several field profiles, $\beta =
  10^{-3}$, $\gamma = 6$ (dashed lines), $\beta = 10^{-1}$, $\gamma =
  4$ (dashed-dotted lines), and $\beta \to \infty$, \ie $B_c =$
  constant (dashed-double-dotted lines).  For each pair of curves the upper
  branch is for $P_\perp$ and the lower branch for $P_\parallel$.
  {\it Right panel} (b):  Dependence of $P_\parallel$ on
  the normalized baryon number density for different magnetic field
  profiles and $B_c=10^{18}$ G. The dots show the reference case
  $B_c=0$.  The solid and dashed lines correspond to $\beta=0.1$ and
  $0.001$, respectively. For each $\beta$ we choose a pair of
  $\gamma$'s; in the first case we have $\gamma =1$ and $\gamma = 4$,
  whereas in the second case $\gamma = 1$ and $\gamma = 6$.}
\label{hyp}
\end{center}
\end{figure}

Fig. \ref{hyp}a shows the EoS of hyperonic matter in the cases of no
magnetic field, constant field as well as for fields with various
density profiles. For nonzero magnetic fields the pressure splits into
parallel and perpendicular components showing anisotropy which arises
from the magnetic field contribution.  From Fig. \ref{hyp}a it is also
evident that there is an onset of instability for some field profiles
due to $P_\parallel$ component of the pressure. This is shown in
Fig. \ref{hyp}b in a more systematic manner. It is clear that for a
given value of $\beta$ the corresponding EoS becomes softer as
$\gamma$ is increased. Consequently, beyond a certain critical value
of $\gamma$ and in a certain density regime $P_\parallel$ ceases to
increase and subsequently decreases with  further increase in
$n_b$. This implies that matter becomes unstable above that value of
density for that particular $B_c$ and magnetic field profile. For
comparison we also show results for each $\beta$ with the minimum
value of $\gamma$ taken to be $1$. Note that the maximum value of
$\gamma$ is taken such that $P_\parallel$ forms a plateau as a
function of $n_b$.

The instability arises due to the negative contribution from the field
energy density (pressure) to the net pressure of magnetized matter in
the direction of the magnetic field. Since for any particular $B_c$
and magnetic profile the field strength increases with the increase
of $n_b$ more negative contribution is added to $P_\parallel$ with
the increase of $n_b$. Consequently, at a certain density,
$P_\parallel$ ceases to increase and then decreases with the increase
of $n_b$.

\begin{figure}
\begin{center}
\includegraphics[width=7cm,height=6cm]{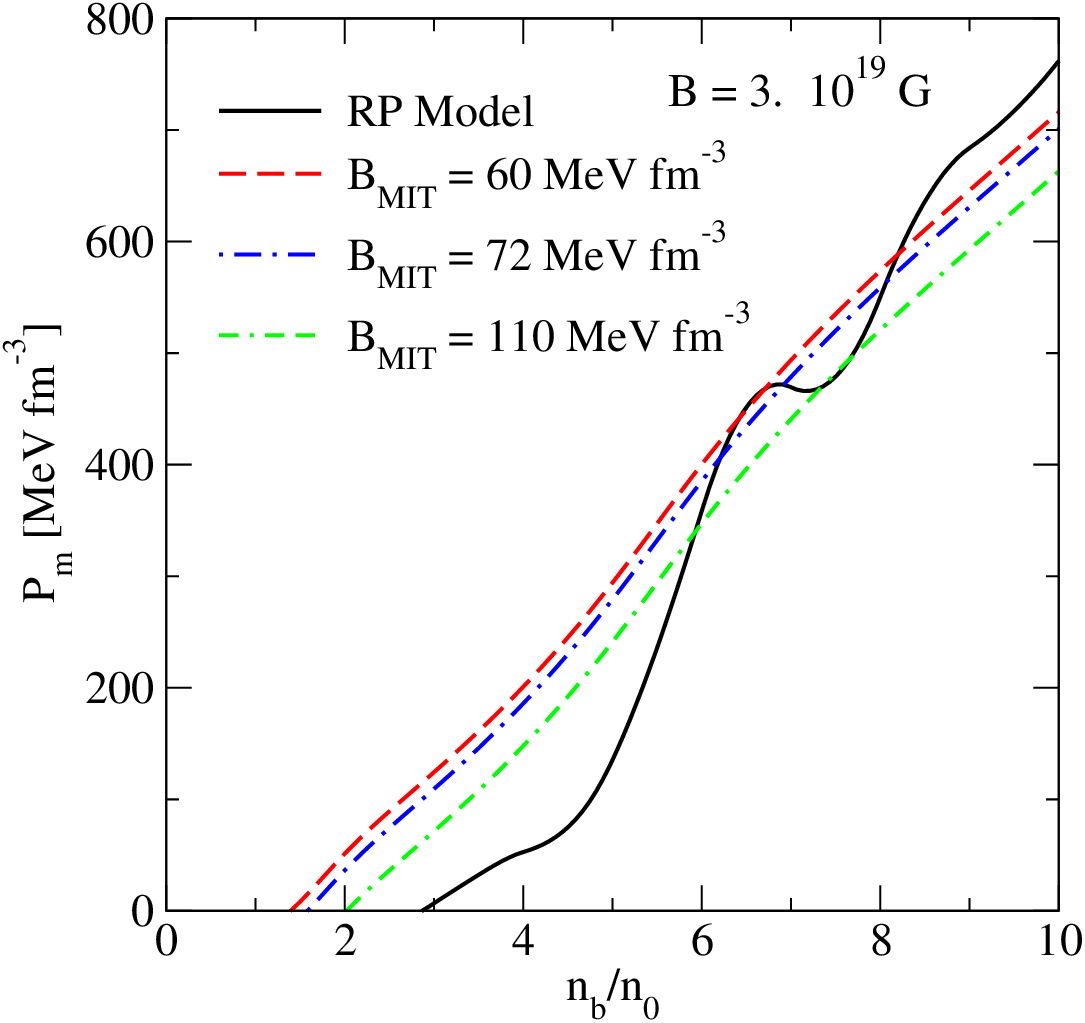}
\hskip 1.0cm
\includegraphics[width=7cm,height=6cm]{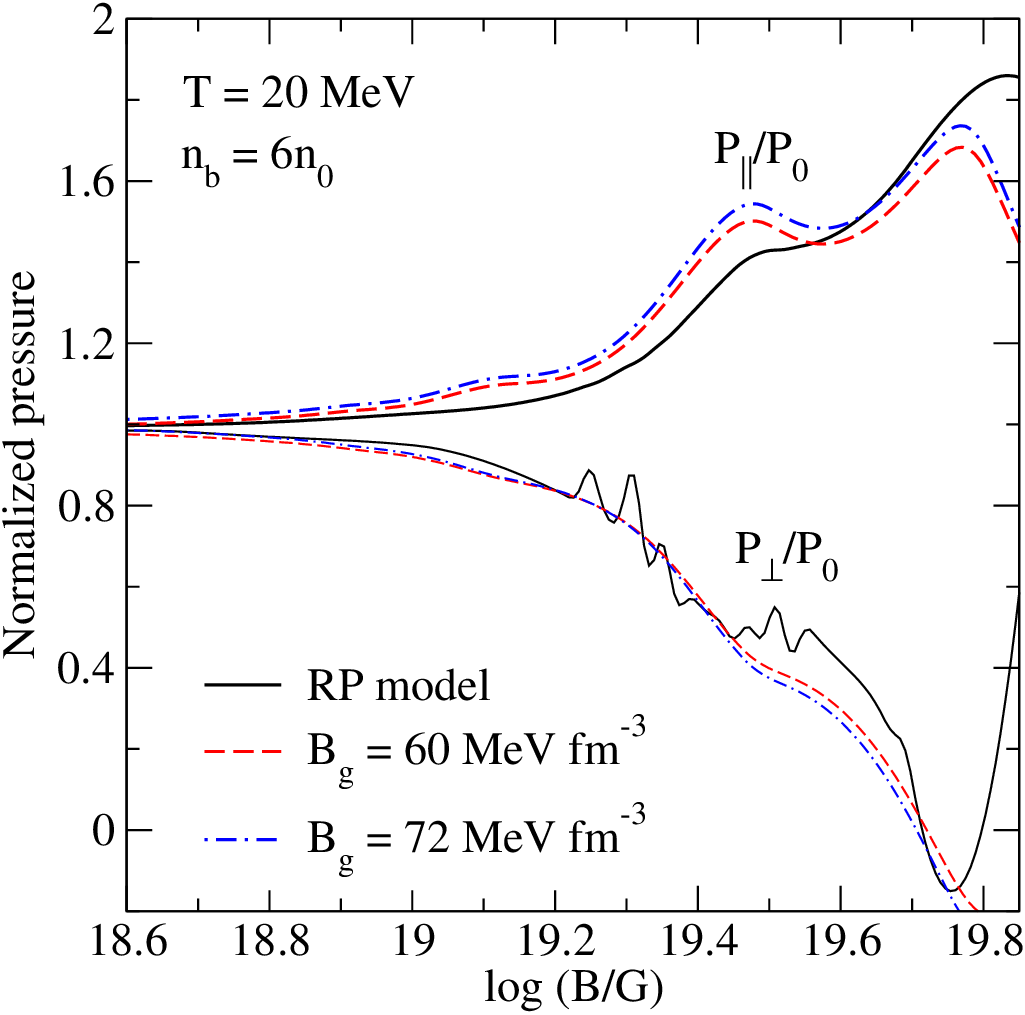}
\caption{ {\it Left panel} (a): Dependence of the thermodynamic
  pressure $P_m$ on the normalized baryon number density at $T = 20$
  MeV for $B =3\times 10^{19}$ G.  The pressure is shown for the RP
  model (solid), for the MIT bag model with $B_{\rm MIT} = 60$ MeV
  fm$^{-3}$ (dashed), $B_{\rm MIT} = 72$ MeV fm$^{-3}$ (dash-dotted)
  and $B_{\rm MIT} = 110$ MeV fm$^{-3}$ (double-dash-dotted).  {\it
    Right panel} (b): Dependence of the normalized pressure components
  that are parallel and perpendicular to the magnetic field on the
  strength of the magnetic field at $n_b=6n_0$ and $T=20$ MeV for
  various models. The upper three curves correspond to the parallel
  pressure, the lower three - to the perpendicular pressure.  }
\label{sqm}
\end{center}
\end{figure}

We turn now to the problem of quark matter in strong magnetic fields
and we focus below on the effect of the RP model on the magnetized SQM
at finite temperature. Our investigation shows that the results are
not sensitive to the temperature in the range relevant to the physics
of compact stars. Hence, we report here results with a fixed
temperature $T=20$ MeV. In Fig.~\ref{sqm}a we show the EoS of SQM for
the RP model and the bag model. Magnetic field introduces some
oscillations in the pressure with density; in each case the increase
of pressure after a plateau is caused by the opening of a new Landau
level.  The oscillations are much stronger in the RP model and this
can be traced back to the momentum dependence of the potential. The
major contribution comes from the static gluon propagator part of the
potential, while the logarithmic factor in the potential depends
weakly on momentum. Note that at a certain density the pressure has a
plateau and slight negative downturn, which can be interpreted as an
instability of homogenous magnetized matter towards phase separation.

For large magnetic fields, the effects of anisotropy become important,
in particular the parallel and perpendicular components of pressure
differ substantially. We show the variations of $P_\parallel$ and
$P_\perp$ with $B$ within the RP and bag models at $n=6n_0$ in
Fig.~\ref{sqm}b. We note that below $B=3\times10^{18}$~G, both
$P_\parallel$ and $P_\perp$ are practically equal to the pressure of
matter in absence of magnetic field. Hence, we conclude that for the
SQM the effect of magnetic field is not significant below $B\sim
10^{18}$~G in the framework of our current modes. For larger fields
$P_\parallel$ increases whereas $P_\perp$ decreases in both models of
SQM. For large enough fields $P_\perp$ becomes negative starting from
some critical value of $B$; this critical value is almost the same in
both models. We now recall that for very large magnetic field
$P_\perp\rightarrow 0$ in the models without
confinement~\cite{2010PhRvD..81d5015H,2012EPJA...48..189D}.  We see
that the addition of the confining potential provides additional
``attraction" inside the SQM and its effect becomes more transparent
at larger $B$.

\section{Conclusions}\label{conc}

We have found that above a certain critical field value $B_{cr}\sim
10^{18}$ G the hyperonic matter may become unstable.  The instability
arises due to negative contribution of the field pressure to the
pressure of matter. The details of the onset of instability depend on
the assumed central field value $B_c$ as well as the parameterization
of the field profile. The instability puts a natural upper bound on
the possible central magnetic field of a neutron star.

We find significant differences between the equation of state of the
strange quark matter predicted by the RP and the bag models. This is
due to the intrinsic momentum-dependent interaction between quarks in
the RP model, which mimics the one-gluon-exchange interaction of the
QCD. Specifically, we find that the thermodynamic pressure in the RP
model is more sensitive to baryon density when the magnetic field is
strong. The de Haas-van Alfven type oscillations in the transverse
pressure $P_\perp$ are much more pronounced in the RP model than in
the MIT bag model.

Furthermore, we find that the presence of a confining potential,
modeled either in terms of the Richardson potential or the MIT bag,
suppresses the pressure components $P_\parallel$ and $P_\perp$ and, at
large $B$, the anisotropy in the equation of state. The splitting
between the longitudinal pressure $P_\parallel$ and the transverse
pressure $P_\perp$ was found to be weaker than that in free
(non-interacting) strange quark matter. This underlines the importance
of taking into account the confining potential in studies of strongly
magnetic strange quark matter in cores of neutron stars and in strange
stars. It remains an interesting task to explore the effects of the
confining potential on the structure and geometry of strongly
magnetized stars.

\ack
The author acknowledges the support of the Alexander von Humboldt
Foundation.

\end{document}